# Molecular dynamics simulation of nanocolloidal amorphous silica particles
# Part I

S. Jenkins and S.R. Kirk,
*Dept. of Technology, Mathematics & CS,*
*University West, P.O. Box 957, Trollhättan, SE 461 29, Sweden.*

M. Persson and J. Carlen,
*R&D Pulp and Paper, Eka Chemicals (Akzo Nobel) AB, Rollsbo, Sweden.*

Z. Abbas
*Department of Chemistry, Göteborg University, SE-412 96 Gothenburg Sweden.*

**Abstract:**

Explicit molecular dynamics simulations were applied to a pair of amorphous silica nanoparticles in aqueous solution, of diameter 4.4 nm with four different background electrolyte concentrations, to extract the mean force acting between the pair of silica nanoparticles. Dependences of the interparticle forces with separation and the background electrolyte concentration were demonstrated. The nature of the interaction of the counter-ions with charged silica surface sites (deprotonated silanols) was investigated. A 'patchy' double layer of adsorbed sodium counter-ions. was observed. Dependences of the interparticle potential of mean force with separation and the background electrolyte concentration were demonstrated. Direct evidence of the solvation forces is presented in terms of changes of the water ordering at the surfaces of the isolated and double nanoparticles. The nature of the interaction of the counter-ions with charged silica surface sites (deprotonated silanols) was investigated in terms of quantifying the effects of the number of water molecules separately inside each of the pair of nanoparticles by defining an impermeability measure. A direct correlation was found between impermeability (related to the silica surface 'hairiness') and the disruption of water ordering. Differences in the impermeability between the two nanoparticles are attributed to differences in the calculated electric dipole moment.

# I. Introduction

In recent years there has been a rapid increase in the number of technological applications of silica in colloidal form. In addition to the more traditional range of applications (e.g. in the food, paint, coatings, and paper industries), new ranges of applications in, e.g. the biomedical industries [1, 2] have been developed. In these applications, controlling the stability of the silica particles in (usually) aequous solution, the size dispersion of the particles and the interactions between both the particles themselves and between the particles and other chemical species, such as solvent molecules, counter-ions, polyelectrolytes and other functional species, are all of prime importance. It is therefore essential that the physical and chemical phenomena underlying the production of silica colloids with desirable properties be thoroughly understood.

Silica in colloidal form [3] has been used in industrial applications and studied for many decades, both in terms of production techniques [4, 5] and customization of properties. Theoretical models for property prediction have been developed covering a wide range of length and time scales, Ab-initio studies of nucleation [6, 7] and stability of oligomers [8-11] have been carried out; molecular dynamics using reactive potentials has also been used in similar studies [12-15]. In recent years, some multiscale studies on silica systems have also been carried out [16-18]. A number of mesoscale approaches to colloidal structure and dynamics such as Monte-Carlo, Brownian dynamics etc.[19-24] have been in use for a much longer time - these necessarily neglect atomic-level details in order to allow longer simulated length and time scales.

The Derjaguin-Landau-Verwey-Overbeek (DLVO) theory of interparticle interactions [25-27] has been widely and successfully used to predict stability in colloidal systems, in terms of a balance of attractive van der Waals forces and repulsive electrical double-layer forces. While useful in a wide range of systems, this theory does not account for specific ion effects [28], and neglects short- and long-ranged contributions (e.g. dispersion forces [29]) to the interparticle force which are important in some systems. Colloidal silica is such a system, and 'non-DLVO' forces in silica colloids have also been attributed to additional solvation forces and a 'hairy' nanoparticle surface, among others [30-34]. Additionally, DLVO theory does not provide correct predictions of stability and aggregation of silica particles up to sub-micrometer sizes in high background salt and/or low pH environments [35, 36].

The purpose of this study is therefore to complement existing work by investigating the effects of a number of experimental parameters on effective forces and solvent and ion effects between amorphous silica nanocolloidal particles in aequous solution by means of explicit molecular dynamics simulations.

# II. METHODS

**Modelling of amorphous silica particle structure**

An amorphous silica 'starting' particle (approximately spherical, diameter 84 Å, containing 3784 atoms) was created [37] by melting, then quenching a sample of α-quartz within a simulation performed using the DMol3 code [38]. Approximately spherical smaller particles were then created from copies of the 'starting' particle by removing all atoms which lay outside progressively smaller spheres [39] each sphere centered on the 'starting' particle geometrical center, the surface of each 'raw' smaller particle then being repaired and all resultant surface hydroxyl groups reconstructed. The set of progressively smaller test particles generated in this way were equilibrated at 300K before later use.

**Allocation of surface charge on silica test particles**

Negative partial charges were allocated to the test silica particles by deprotonating surface hydroxyl (Si-OH → Si-O$^-$) groups, and assigning an additional -1 partial charge to the deprotonated oxygen. The number of deprotonated sites was determined by the experimental silicon to sodium (Si:Na$^+$) ratio used in the industrial production process (usually in the range 5:1-40:1; for this work we chose a value for the ratio of 20:1). The number of silicon atoms in each particle then determined the number of residual sodium atoms. The number of surface hydroxyl groups to be deprotonated was then set equal to the number of residual Na$^+$ ions, for overall charge neutrality. Deprotonation sites on each test particle were allocated using an Monte Carlo optimization algorithm, taking account of the mutual repulsion of the deprotonated sites. The position coordinates of all surface hydroxyl oxygen atoms were recorded, then a subset of size $N$ of these sites chosen at random.

The sum $S$ over the $N$ sites of $|\mathbf{r_{ij}}|^2$, where $\mathbf{r}_{ij}$ is the position vector between sites $i$ and $j$, was then calculated (avoiding the double sum) for the current subset. One site is then chosen randomly from the subset, swapped with a random site not currently in the subset, and *S* recalculated. If *S* increases as a result of the swap, the swap is allowed, otherwise the swap is rejected. This process is repeated over many steps, with the subset of sites evolving towards the 'maximally repulsive' configuration. The simple form used for the 'fitness function' *S* was chosen for computational efficiency, and to allow for the surface roughness and non-sphericity of the silica particles.

**Molecular dynamics**

Molecular dynamics calculations were carried out in the NVT ensemble using the GROMACS [40-42] code (version 3.3.1), using OPLS-AA [43] force fields, with additional parameters for silica from literature [44]. The water model used was a flexible variant of the TIP4P 4-site model [45, 46]. During actual MD runs, the simulation temperature of all species was set to 300K, particle velocities being chosen from a Maxwell distribution, using a Berendsen [47] thermostat with a coupling time constant of 0.1 ps. Neighbour list, Coulomb and van der Waals cutoffs were all set to 1 nm. Centre of mass motion was removed after every timestep. Long-range electrostatic forces were treated using a particle-mesh Ewald [48] treatment, with long-range dispersion corrections applied. The FFT grid spacing was in all cases around 0.118 nm, varying slightly with simulation box size, and cubic (order 4) spline interpolation was used on the FFT grid. In all cases, a 3-stage MD protocol was used - energy minimization, followed by 100 ps (0.002 ps time-step) of position-restrained MD, followed by the 'production' run (performed with 0.001 ps time-step). In cases where a non-zero background salt concentration was modelled, an appropriate number of water molecules in the simulation box were randomly substituted with equal numbers of $Na^+$ and $Cl^-$ ions to reach the desired concentration.

**Single-particle MD runs**

The test nanoparticles were centered in cubic simulation boxes with periodic boundary conditions; the boxes were then filled with water, the box size being chosen so that there was at least 1 nm of water between any side of the box and the nanoparticle, in order to avoid potential spurious water structuring effects caused by the periodicity of the simulation box [49]. Sodium ions were substituted in randomly for water molecules until the contents of the simulation box was overall electrically neutral. The production run in these cases consisted of a 500 ps unconstrained MD run, all other parameters being as previously stated.

**Double-particle MD runs - PMF calculations**

The fully equilibrated test nanoparticles from the single-particle MD runs were used as building blocks to investigate interparticle interactions. A pair of nanoparticles (along with their accompanying cloud of neutralizing $Na^+$ ions) were placed relative to each other in such a way that the distance between the centres of mass (COM distance) took a number of specified values. The resulting pair of nanoparticles was then centered in a large simulation box with periodic boundary conditions applied. The dimensions of this large simulation box was chosen such that the any part of one nanoparticle was closer to all parts of the other member of the pair than to any periodic copies of either particle. The remaining space in the simulation box was filled with TIP4P water. In these runs the production phase consisted of a 500 ps (0.001 ps time-step) potential of mean force (PMF) calculation, using the constraints method built into the pull code within GROMACS. The COM distance between the nanoparticle pair at the end of the position-restrained MD run was used as a constraint distance. An additional constraint force was applied by the GROMACS code during the PMF production run to maintain this original COM distance, and the value of the constraint force needed to maintain this distance was monitored and recorded at every time-step of the production run. After the end of the production run, the 500,000 values of these constraint forces were averaged, representing the average interparticle attractive or repulsive force caused by interactions between the particles [50]. Variants of these basic PMF runs were generated and carried out for various different background concentrations of NaCl in the solvent (the additional ions having been present from the beginning of the protocol), in the form of additional $Na^+$ and $Cl^-$ ions randomly substituted for solvent (water) molecules. Numbers of additional ions were chosen to replicate molecular background salt concentrations of 0.00M, 0.01M, 0.1M and 1.0M, encompassing the range of background salt levels used in industrial production of these nanoparticles, and found in their common applications.

**III. Results and discussion**

The pair of nanoparticles that form the basis of this study range both have diameters of 4.4 nm (as seen Figure 1(a)). The amorphous silica nanoparticles consist mainly of tetrahedrally coordinated silicon atoms, including silanol bridges, surface hydroxyl groups and deprotonated oxygen surface ($O^-$) sites. The number of surface sites is determined by the Si:$Na^+$ ratio, which was 20:1 in all cases in this work. The number of charged oxygen surface sites (also equal to the number of sodium counter ions) was 49 on each nanoparticle. From Figure 1(b) it can be seen that the sodium ion does not form a strong chemical bond with the oxygen surface site, since the length of $Na^+$--$O^-$ interaction is typically around 2.3 Å, which would instead indicate the formation of a weaker i.e. purely electrostatic (closed shell) interaction. The four water molecules are positioned in the figure to clearly indicate that they are hydrogen-bonded to the surface $O^-$ site. All such observed deprotonated surface sites showed this characteristic. The lack of strong chemical bonding of the sodium (or surface adsorption) to this site has important consequences for the aggregation properties of the silica nanoparticles. The calculated average cosine of the water orientation angle for an entire MD trajectory (between the line joining the surface oxygen site to the water O atom, and the in-plane bisector of the water molecule) around the surface oxygen always has a value of approximately -0.6, for both the single-particle and double-particle cases. This observation provide strong evidence of local water ordering around the deprotonated surface sites, possibly partially induced by the effect of the nearby accompanying sodium counter-ion [32]. The observed average counter-ion-surface site separation is around 2.3 Å (c.f. the radius of a hydrated $Na^+$ ion, 2.37 Å [51]); by contrast, the hydrogen bond distance from the $Na^+$ counter-ion to the neighbouring water molecules is around 2.6 Å, again demonstrating the strong effect of the presence of the silica surface on the local water ordering. Figure 2 shows the axial-radial distribution of sodium ions around the silica nanoparticles. The classic 'double-layer' can be observed to be 'patchy' rather than continuous, despite the fact that the results pertain to the entire MD trajectory rather that a static 'snapshot', even with a significant number of deprotonated surface sites. Figure 3(a-d) show the variation of the mean interparticle force with COM (centre-of-mass) separation, along with running averages taken over 10 data points (shown as dotted lines). Table 1 also shows a summary of the important features of the inter-particle forces. The sub-figures (a-d), showing in each case the effects of increasing amounts of background sodium.

For simple spherical large particles with smooth hard surfaces, and separated by small spherical solvent molecules, one would expect that the solvation forces would obey a decaying oscillatory function with particle separation, including both the attractive depletion and repulsive structural energy barrier [52-54]. In our case, however, the silica nanoparticles are neither perfectly spherical, nor smooth, nor hard, so it is not surprising that their inter-particle forces do not show smooth oscillations. The addition of background sodium for the pair of nanoparticles has the effect of increasing the initial repulsive barrier, up to a maximum for a background sodium concentration of 0.10M, as can be seen from Figure 3(a-d). Notice in particular that the variation of the inter-particle force for the nanoparticles follows a oscillatory repulsive form, suggestive of hydrophilic surface character. This effect may be due to overcharging, due to the number of sodium ions not being 'accommodated' on the surface sites fully. This is also evident in Table 1 from the entries for background $Na^+$ concentration of 1.00M, where the average forces for these particles are rather repulsive. This is also true at a background $Na^+$ concentration of 0.01M. Notice from Table 1 the rather large attractive forces. These large attractive forces may well be due to the relatively large surface area of these particles compared to the number of deprotonated sites (49) (which can accommodate more sodium ions before they start repelling each other, since the surface area increases proportional to $r^2$ and the number of surface sites as $r^3$; the number of sites is determined by the silicon to sodium ratio; a constant for this work). The magnitudes of the minimum and maximum inter-particle forces given in Table 1. The magnitude of the interparticle forces tend to zero with increasing particle separation, except for the case of 0.00M background $Na^+$ concentration. This is to due to the nanoparticles particles not being separated far enough for the available $Na^+$ ions to screen electrostatic forces, this situation could be remedied by in future ensuring that the separations are increased until the interparticle forces tend to zero with increasing particle separation.

To summarise: the average force for a given sodium concentration gives an indication of the particles ability to accommodate the sodium ions. The average force does not vary significantly with increasing $Na^+$ concentration. There is an increase in the maximum repulsive force with increasing background sodium concentration from 0.01M to 0.10M but a decrease at the highest sodium concentration.

In Figure 4 we show the calculated potential of mean forces (PMF) for the pair of particles (4.4 nm in diameter) used in this study, by integration of the mean interparticle forces. We removed the constant of integration by estimating the potential at infinite range (i.e. zero) to be the value at the largest range used in our calculations. In general, we observed that the multiple simulation runs started with initial COM distance values chosen with a constant increment (0.5 nm) yielded constraint distances for the PMF phase of the simulation runs (i.e. after the preliminary temperature equilibration position-restraints phase) which did not have a constant spacing, as can be seen from Figure 4, under the influence of the oscillatory interparticle forces. Notice that the form of a depletion well and a repulsive structural barrier stabilizer [52-54] can be seen; this barrier contributes in preventing large particles from flocculating or coalescing.

The hydrogen bond length (hydrogen-acceptor oxygen) distributions between water molecules (see Figure 5(a)) was determined from the PMF MD trajectories, the peak in the distribution being found at 0.18 nm. This compares well with the experimentally determined hydrogen bond length in water [55]. The distribution of hydrogen bond lengths between nanoparticles and water molecules is shown in Figure 5(b); the position of the first peak being found at 0.15 nm and the second at 0.25 nm. From Figure 5(c), the peak in the length distribution of the hydrogen bonding between the surface oxygen sites $O^-$ and water molecules was located at a distance of 0.15 nm; this shows the hydrogen bonds are rather short due to the excess charge at the oxygen surface site. Note also that this peak coincides with the first peak in Figure 5(b).

In order to quantify the spatial effects of the various background concentrations of $Na^+$ on the surface and beyond we have calculated a set of three radial distribution functions (*rdf*s); the *rdf*s for water around the sodium ions and the deprotonated oxygen surface sites ($O^-$) and for the sodium ions around the deprotonated oxygen surface sites ($O^-$), shown in Figures 6(a-c) respectively. Examining the figures in more detail reveals some important details about the nature of the surface of the silica nanoparticles. In Figure 6(b) the first peak is at a distance of 0.15 nm from the $O^-$ site, the second at 0.24 nm. When this is compared with Figure 6(c), where the first peak is at 0.23 nm and the second is at 0.35 nm, it is apparent that the water molecules are closer to the $O^-$ surface site than are the sodium ions. Figure 6(b) also reflects the results of the hydrogen bonding distributions in Figure 5(b). Another interesting feature is the fact that in both Figure 6(a) and (b) the *rdf* after the first peak is close to zero. This is more similar to the typical *rdf* of a solid rather than a liquid, showing as it does some ordering or layering effects in the water around the surface of the nanoparticles.

In Table 2 a summary of the *rdf* data is presented for the pair of silica nanoparticles considered in this study, listing both the pairwise interaction of the nanoparticles and the isolated nanoparticle for comparison. Examining the first set of entries ($Na^+$ and $H_2O$) in Table 2 it can be seen that there is a monotonic increase of the height ratio (second peak to first) with background sodium concentration. This variable is used to quantify the extent to which the second species is 'piling up' around the first species used in the analysis. This holds true for the double particles and isolated particles. The change in the $Na^+$ and $H_2O$ *rdf* ratio with increasing background sodium concentration could be explained in terms of the sodium ions binding the water and preventing it from entering the nanoparticle.

The time-averaged number of water molecules found within the nanoparticles was extracted from the PMF MD trajectories (as shown in Figure 7(a-d), and the mean values of these numbers across all investigated interparticle separations listed in Table 3. The relationship between the background sodium concentration and the amount of water entering the nanoparticle is not simple (see Table 3), as the instance of 0.00M background sodium does not correspond to the largest number of water molecules entering the nanoparticle, which could be due to the surface density of charged surface sites, and hence associated sodium counter-ions repelling the background $Na^+$ as the background $Na^+$ concentration is increased. In addition, the isolated particle case in Table 3 has more water molecules entering the nanoparticle in all background sodium concentrations as compared with any of the sodium background concentrations for the double nanoparticle case. This is evidence that the sodium double layer of counter-ions is acting more strongly in the double particle than the isolated particle, as seen by the fewer sodiums entering each of the pair of the double particle systems (see Table 3). Evidence that the double layer of sodium counter-ions is sufficient to attract water molecules, is shown in Figures 8(a-b) (where the plots show the of variation of the number of water molecules with background sodium concentration within 0.5 nm of the O- surface).

The solvation forces (also referred to as hydration forces for water) do not occur because the water molecules have formed semi-ordered layers at the surface, but rather because of the disruption (or changes) of the ordering due to the approach of a second surface [52-54]. This can be seen if we compare the orientation plot (as shown in Figure 9(a) and 9(b) - the definition of the orientation can be found in the Figure caption) of the water molecules around the isolated particle and the double particle respectively. Notice in particular, that the difference in these Figures (see accompanying figure captions also) in the vicinity of 2 nm. From Table 3, it can be seen from the low values of the impermeability (defined as the dry core radius of a nanoparticle measured radially from the centre of mass of the nanoparticle, in nanometres) that the nanoparticle is rather 'hairy'. This observation has important consequences for explaining the lack of regular oscillatory behaviour of the mean force vs. separation plots. It is worth noting that the level of 'hairiness' can be controlled within an industrial setting. The increased 'hairiness' of the 4.4 nm nanoparticle, demonstrated by the low values of the impermeability, will disrupt the water ordering at the surface and hence disrupt the oscillatory form of the forces, making particle aggregation more difficult than would be the case for a less hairy particle.

In Table 2 for the entry $O^-$ and $H_2O$, notice in particular that for the isolated particle the *rdf* ratios for 0.00M and 1.00M background $Na^+$ are nearly identical. This is not the case for the double particle. For the final *rdf* ratio, that is $O^-$ and $Na^+$ in Table 2, the double particle and isolated particle cases have similar values for a given background $Na^+$ concentration, the exception being for 0.00 M background $Na^+$. The small values of the $O^-$ and $Na^+$ *rdf* ratios show that the sodium ions are rather more confined around the silica nanoparticles, i.e. at the first peak in the *rdf*.

In Table 3, clear differences emerge; for the double particle case, nanoparticle 1 (denoted 'snp 1') has a higher value of the impermeability than nanoparticle 2 (denoted 'snp 2') for all investigated values of the background sodium concentration. In addition, the isolated nanoparticle has consistently much lower values of the impermeability than either of the nanoparticles in the double particle case. This is consistent with the isolated nanoparticle having more water molecules penetrate it, and shows that the findings in this work go beyond starting geometries, i.e. since we use one nanoparticle to create the pair of nanoparticles. In effect the isolated nanoparticle is 'softer' than either of the twin nanoparticles. This can be explained in terms of the weaker double layer of sodium counter-ions in the isolated nanoparticle than in the double nanoparticle case. It can be seen that the variation of the number of water molecules inside the pair of nanoparticles does not vary regularly with particle separation; see Figures 7(a-d). There are, however, as has been mentioned earlier, predictable trends in the number of water molecules inside the nanoparticles when time-averages are taken over all background sodium concentrations (see Table 3). Notice that there are variations in the number of water molecules inside each of the pair of nanoparticles - a physical reason for this could be that the two particles have different calculated electric dipole moments. The individual nanoparticles are distinguishable and labelled 'snp 1' and 'snp 2' in Figure 7, and have average (over separation) dipole moments of approximately 31,000 and 42,000 Debye respectively. The isolated nanoparticle had a dipole moment of 24,000 Debye. The value of the dipole moment for the double and isolated nanoparticles is found to be independent of the molarity of the background sodium. These variations in the dipole moments could explain the differences in the number of water molecules trapped inside the two nanoparticles (see Table 3) for a given background sodium concentration.

**Conclusions**

This study has successfully modelled the interactions between, and near-surface structure of, amorphous silica nanoparticles in aqueous solutions with varying amounts of background counter-ions using realistic molecular dynamics force fields, evaluating the interparticle potentials using a PMF formalism. This work shows that water molecules can penetrate into amorphous silica nanocolloid particles, and that 'hairiness' of the silica surface (modelled at atomic resolution with a realistic silica surface structure) has an effect on the interparticle potential of mean force, in agreement with other work in the literature [17,18], describing a thin 'hairy' layer on the surfaces of such systems. The consequences for water ordering has also been investigated, in conjunction with the presence of surface-bound and free background counter-ions. The hydrogen bonding characteristics in the surface region were found to be independent of background $Na^+$ concentration. In addition, we have quantified and examined the factors influencing the effectiveness of the sodium counter-ion 'double layer' in preventing or enabling water molecules entering the silica nanoparticle. Reactive MD potentials [55] and QM/MM [56] treatments may provide a future avenue for investigations of the longer time-scale interactions between 'crashed' particles as condensation reactions between silanol groups create bonds as the particles 'fuse' together. At longer length and time-scales, the interparticle potentials derived in this work will inform ongoing coarse-grained MD investigations of flocculation and gelation in silica nanocolloid systems, with the eventual goal of creating a microscopic theory of gelation in partnership with novel application of mathematical network formalism to utilize the universal structural character of colloids [57].

Work is also currently underway to determine the quantum mechanical details of the interactions between the charged surface oxygen sites and the accompanying sodium counter-ions, using the Theory of Atoms in Molecules [58] within the QM/MM approximation [56]. Other further investigations of the effect of altering the Si:$Na^+$ ratio for the nanoparticles, currently underway [59], may yield rather different PMF profiles to the work described here. Additional future work will also involve exploring factors influencing the effect of the form of the PMF and ultimately the aggregation, e.g. by selecting species of counter-ions which chemically bind to the $O^-$ surface sites.

## Acknowledgments

The Knowledge Foundation are gratefully acknowledged for the support of SJ and SRK, grant number 2004/0284. This work was made possible by the facilities of the Shared Hierarchical Academic Research Computing Network (SHARCNET:www.sharcnet.ca), through the kind auspices of our sponsor Dr. P.W. Ayers, Department of Chemistry, McMaster University, Ontario, Canada.

**Table 1.** Variation of the inter-particle force for an entire MD trajectory with separation, and background sodium concentration (see Figure 3(a-d) for plots of the inter-particle forces vs. separation). The first number under each concentration heading is the average force and the maximum values of attraction and repulsion of the inter-particle forces are shown in italicized, and the units used are kJ mol$^{-1}$ nm$^{-1}$.

| nanoparticle-diameter(nm) | Background sodium ion concentration(molarity) | | | |
|---|---|---|---|---|
| | 0.00 | 0.01 | 0.10 | 1.00 |
| 4.40 | 3.797(*-127.8, 94.85*) | -3.362(*-80.54, 106.8*) | 3.620(*-90.19, 110.6*) | 6.214(*-63.83, 70.89*) |

**Table 2.** Summary of the radial distribution function ratios for the silica nanoparticle(s) with diameter 4.4 nm. Each entry in the table is the ratio of the heights of the second to first peaks of the corresponding radial distribution function for each background sodium concentration given by the column heading (see also Figure 6).

| RDF species | Background sodium ion (Na$^+$) molarity | | | |
|---|---|---|---|---|
| | 0.00 | 0.01 | 0.10 | 1.00 |
| **Na$^+$ and H$_2$O** | | | | |
| *snp 1 + snp 2* | 0.390 | 0.406 | 0.465 | 0.473 |
| *isolated* | 0.380 | 0.423 | 0.471 | 0.572 |
| **O$^-$ and H$_2$O** | | | | |
| *snp 1 + snp 2* | 0.440 | 0.446 | 0.448 | 0.427 |
| *isolated* | 0.462 | 0.453 | 0.441 | 0.463 |
| **O$^-$ and Na$^+$** | | | | |
| *snp 1 + snp 2* | 0.047 | 0.040 | 0.050 | 0.061 |
| *isolated* | 0.061 | 0.038 | 0.049 | 0.063 |

**Table 3.** The mean of the time-averaged number of water molecules trapped inside the nanoparticles, (see also Figure 7(a-d)). The individual nanoparticles are distinguishable within the table and Figure 7, labelled 'snp 1' and 'snp 2', and 'isolated' refers to calculations performed on the isolated nanoparticle. The figures in brackets describe the mean of the time-averaged value of the impermeability of the nanoparticles given in nanometres; see Figure 9(a-b) and accompanying text.

**Mean of the time-averaged number of trapped water molecules and impermeability**

| species | Background sodium ion (Na$^+$) molarity | | | |
|---|---|---|---|---|
| | 0.00 | 0.01 | 0.10 | 1.00 |
| *snp 1* | 80.2 (0.468) | 80.5 (0.443) | 79.5 (0.437) | 72.9 (0.475) |
| *snp 2* | 80.9 (0.374) | 81.1 (0.354) | 80.4 (0.379) | 73.5 (0.347) |
| *isolated* | 92.1 (0.252) | 94.7 (0.292) | 93.3 (0.290) | 85.5 (0.253) |

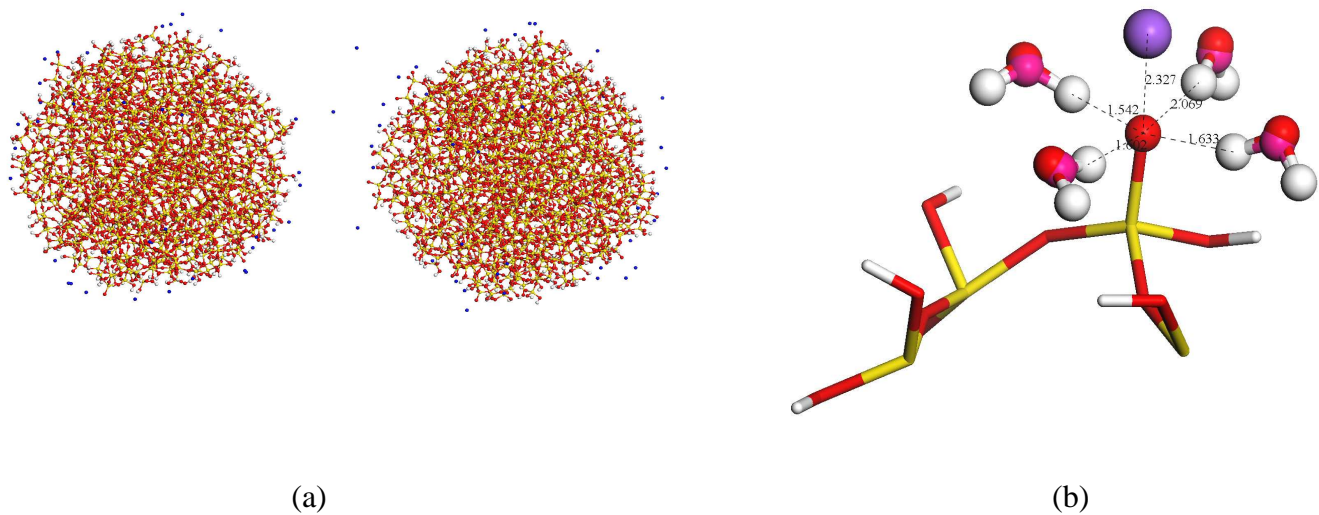

(a) (b)

**Figure 1.** In (a) the pair of nanoparticles with diameter 4.4 nm is shown with the sodium counter ions in blue (visible on the electronic version) and without the surrounding water molecules for clarity. In (b) a snapshot of the local arrangement of the water molecules around a deprotonated oxygen surface site is shown by way of example. It can be seen that four water molecules surround the oxygen surface site; the water hydrogen-surface oxygen separations are shown in Ångstroms. In addition, a sodium counter-ion is shown as a larger sphere at a separation of 2.327 Å.

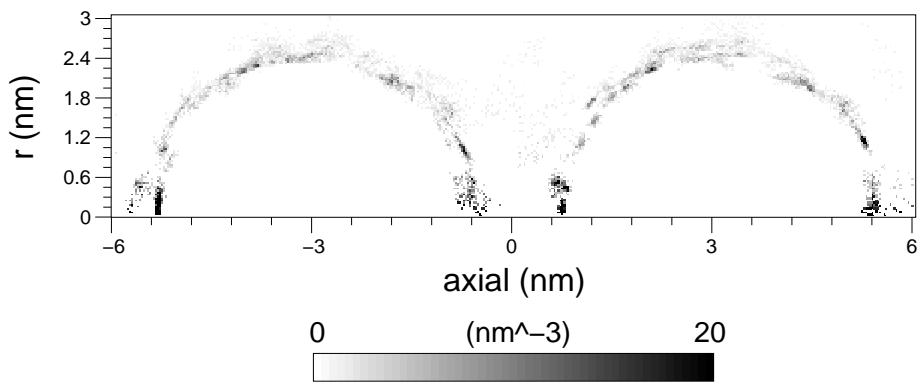

**Figure 2.** The axial-radial density plot of the distribution of sodium counter-ions around the pair of nanoparticles for the entire MD trajectory. The axis corresponds to the line joining the centres of mass of the two particles.

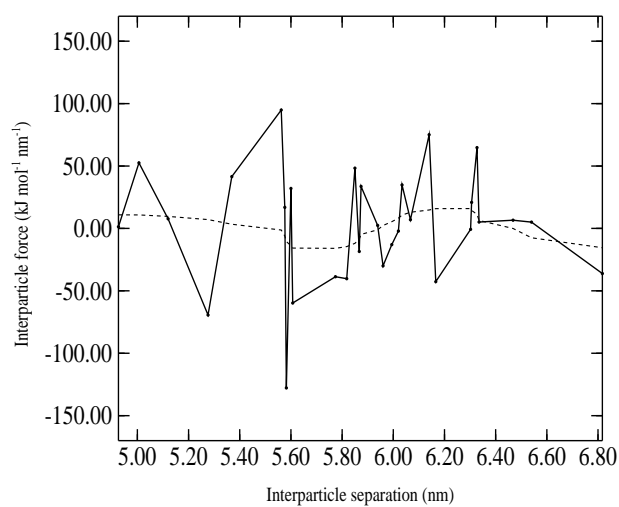 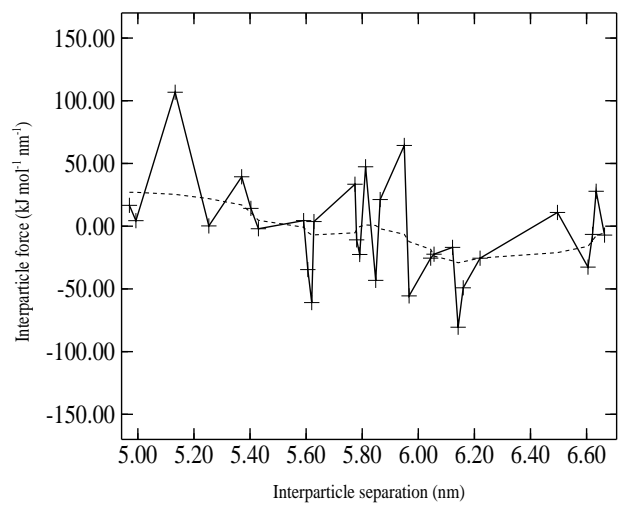

(a) (b)

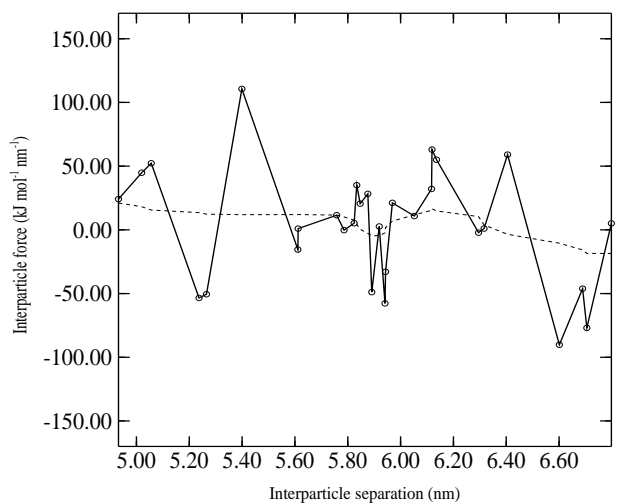
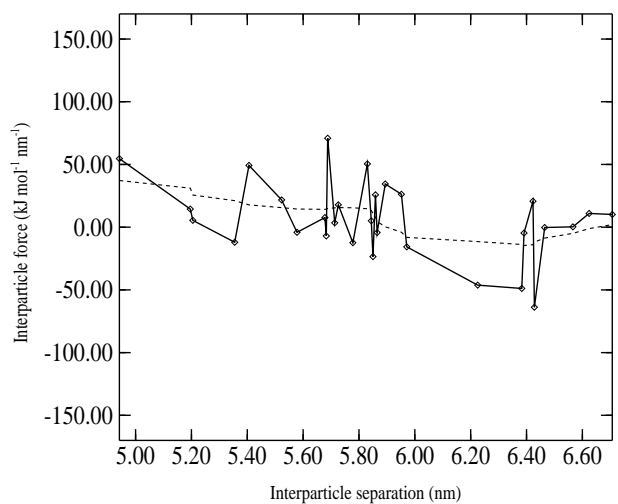

**Figure 3.** Plots of the inter-particle force vs. inter-particle separation for a pair of nanoparticles with diameter of 4.4 nm, are shown in (a)-(d) for background sodium concentrations of 0.00, 0.01, 0.10 and 1.00 M respectively. Overlaid is the running average, shown as a dashed line.

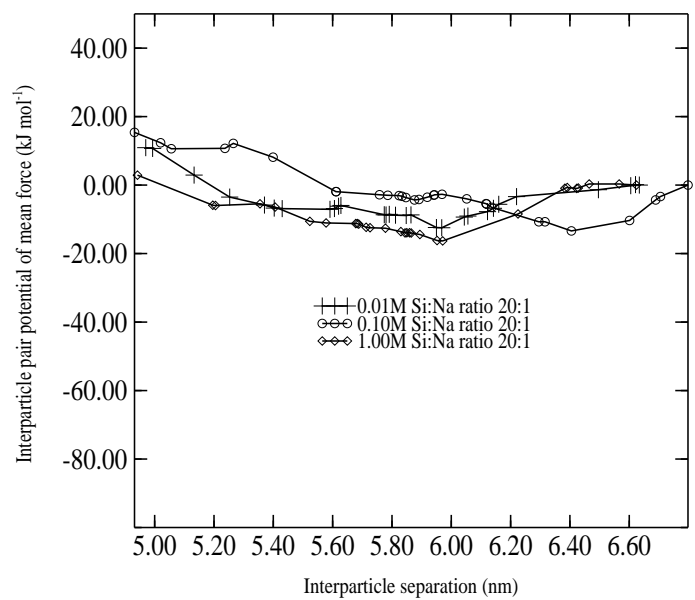

**Figure 4.** The potential of mean forces (PMF) for the pair of particles (with diameter 4.4 nm) in this study shifted to estimate for the infinite range potential.

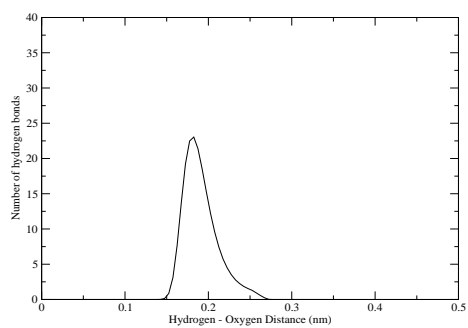
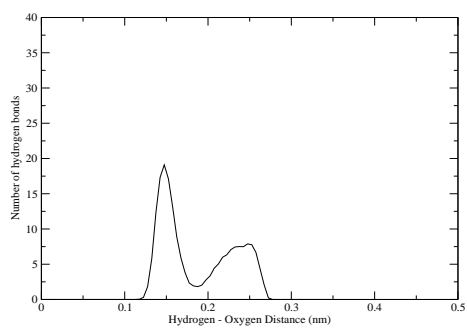
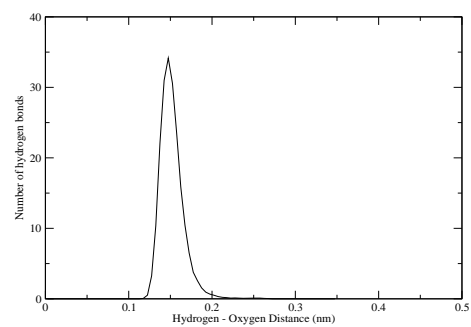

**Figure 5.** (a) The hydrogen bonding length (hydrogen-acceptor) distribution between the (flexible TIP4P) water molecules in the simulation; (b) shows the distribution of hydrogen bond lengths between nanoparticles and water molecules. The peak in the hydrogen bond length distribution of the hydrogen bonding between the surface oxygen sites O$^-$ and water molecules is given in (c).

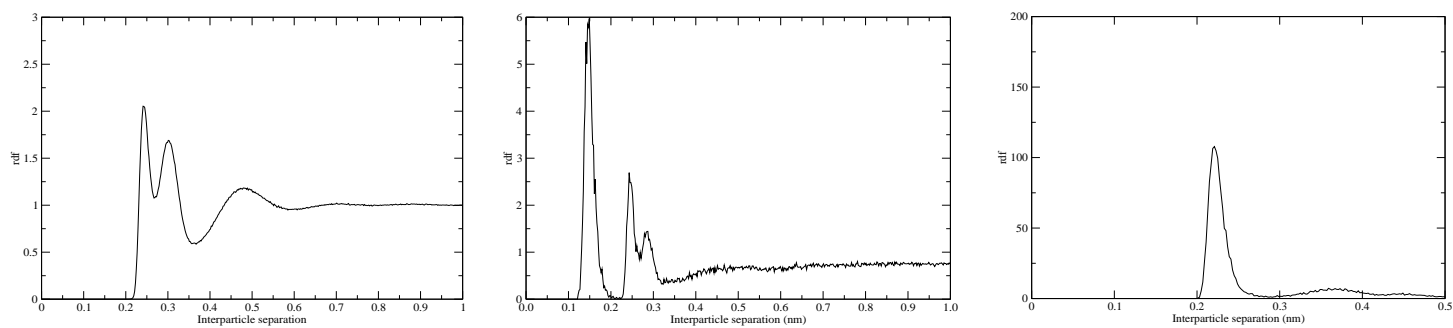

**Figure 6.** The radial distribution functions (*rdf*) for water around the sodium ions is shown in (a), water around the deprotonated oxygen surface sites (O⁻) shown in (b) and the sodium ions around the deprotonated oxygen surface sites (O⁻) is shown in (c).

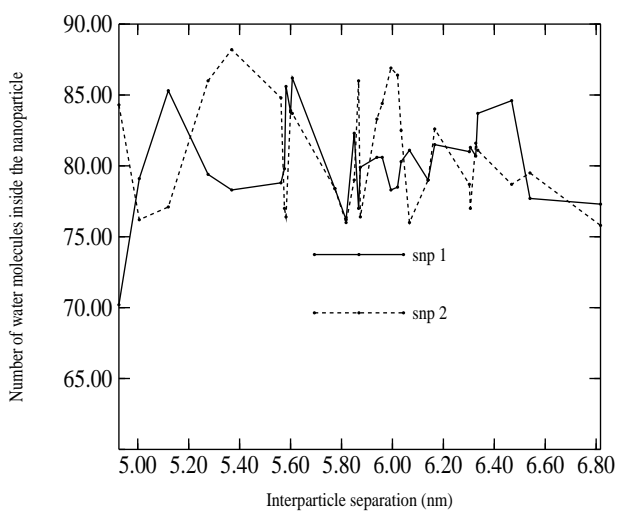

(a)

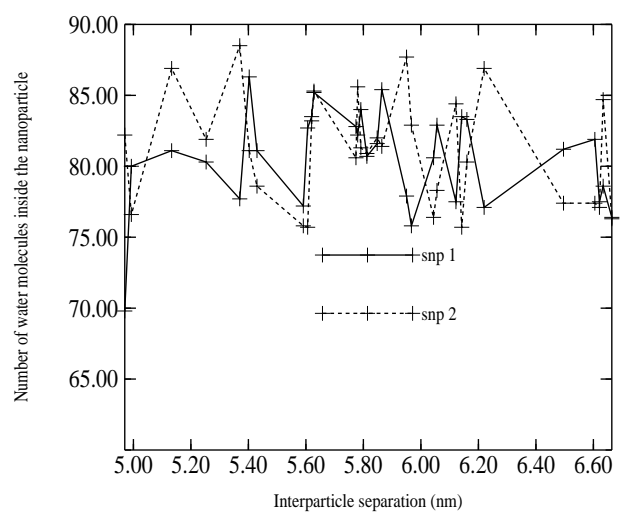

(b)

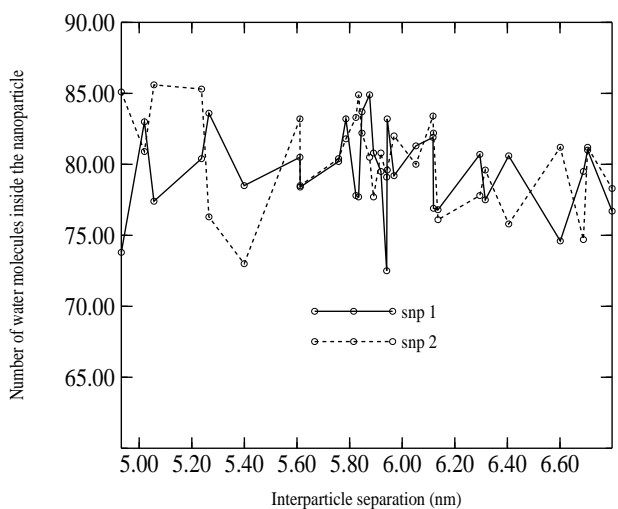

(c)

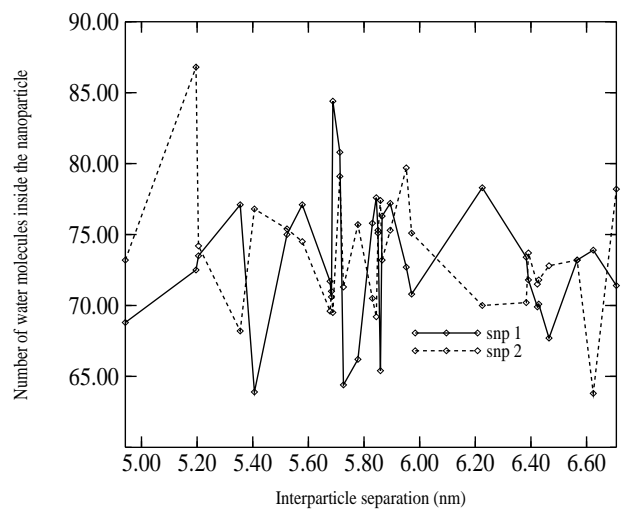

(d)

**Figure 7.** The plots showing the variation of the average number of the water molecules inside the specified radius, with background sodium concentration are shown in (a-d) (see Table 3). The individual nanoparticles are distinguishable within the figure, and labelled 'snp 1' and 'snp 2'.

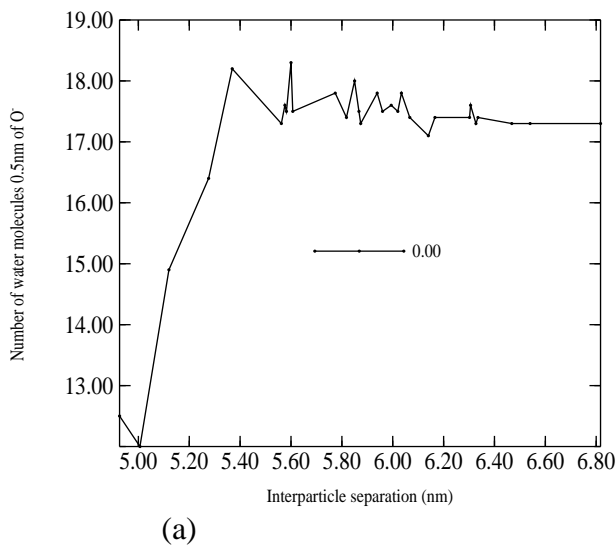 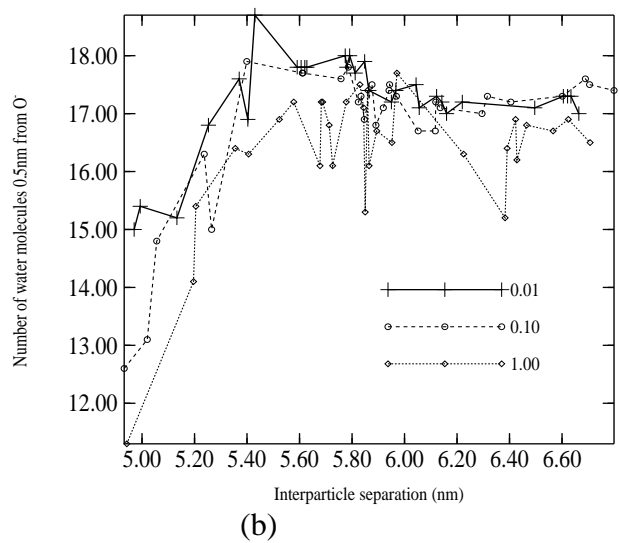

**Figure 8.** The variation of the time-averaged number of water molecules within a 0.5 nm radius of the O⁻ deprotonated silanol surface sites.

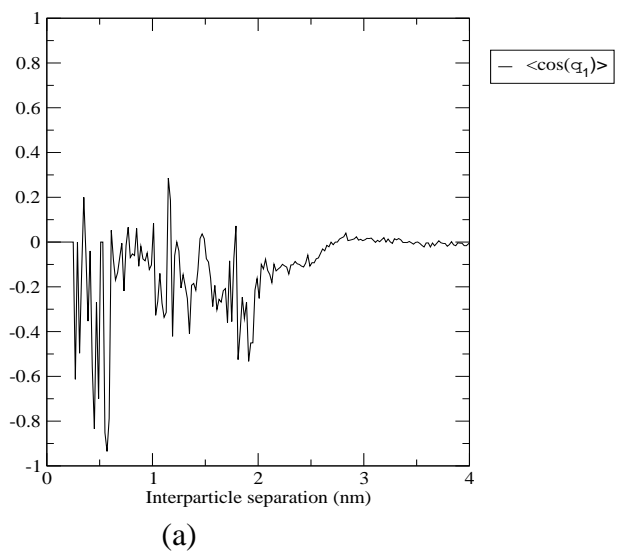 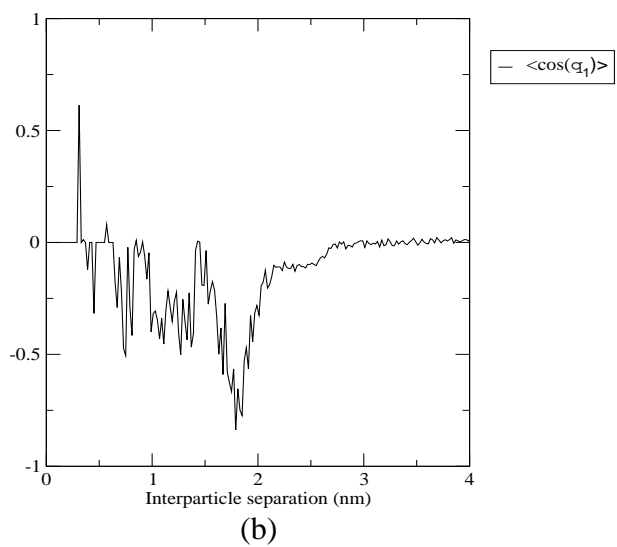

**Figure 9.** The time-averaged orientation of water molecules within the defined radius relative to the centre of mass of a silica nanoparticle shown for the isolated particle in (a) and for one particle from a pair in (b). The outer cutoff radius for the time-averaged water molecule orientation of the nanoparticle is 2.2 nm. A value of +1 or -1 for $\cos(q_1)$ represents a water molecule oriented so that the bisecting in-plane vector (of the water molecule) is parallel or anti-parallel respectively to the vector joining the centre of mass of the nanoparticle to the oxygen atom of the water molecule. The impermeability is given (from the origin along the x-axis) as the maximum extent to which $\langle\cos(q_1)\rangle = 0$ is true, see Table 3.